\documentclass{article}
\usepackage{bm,latexsym,amsmath,amssymb,amsfonts,fancyhdr,color,graphicx,multirow,axodraw}
\usepackage[a4paper,bottom=2cm,top=2.5cm,head=5mm,width=18cm,dvipdfm]{geometry}
\usepackage[usenames,dvipsnames,svgnames,table]{xcolor}
\usepackage[ps2pdf,
            colorlinks=true,
            linkcolor=red,
            urlcolor=blue,
            citecolor=green,          
						bookmarks=true,
						bookmarksnumbered=true,
						breaklinks=true,
						pdfpagemode=Fullscreen,
						pdfstartview=FitBH]{hyperref}
\usepackage[dotinlabels]{titletoc}
\usepackage{titlesec}
\usepackage{authblk}

\numberwithin{equation}{section}
\allowdisplaybreaks[4]

\titlelabel{\thetitle.\quad \hspace{-0.8em}}
\titlecontents{section}
              [1.5em]
              {\vspace{4mm} \large \bf}
              {\contentslabel{1em}}
              {\hspace*{-1em}}
              {\titlerule*[.5pc]{.}\contentspage}
\titlecontents{subsection}
              [3.5em]
              {\vspace{2mm}}
              {\contentslabel{1.8em}}
              {\hspace*{.3em}}
              {\titlerule*[.5pc]{.}\contentspage}
\titlecontents{subsubsection}
              [5.5em]
              {\vspace{2mm}}
              {\contentslabel{2.5em}}
              {\hspace*{.3em}}
              {\titlerule*[.5pc]{.}\contentspage}

\newcommand{\titledef}{JUNO and Neutrinoless Double Beta Decay}

\hypersetup{ pdfauthor = {Shao-Feng Ge},
	     pdftitle = {\titledef}, 
	     pdfsubject = {}, 
             pdfkeywords = {}, 
	     pdfcreator = {LaTeX with hyperref package},
	     pdfproducer = {dvips + ps2pdf} }

\definecolor{gesfpurple}{rgb}{0.47,0.19,0.42}

\definecolor{gesflanse}{rgb}{0.00,0.50,0.50}

\definecolor{gesfblue}{rgb}{0.08,0.42,0.76}

\definecolor{gesfred}{rgb}{1,0,0}

\definecolor{gesfwhite}{rgb}{1,1,1}

\definecolor{gesfblack}{rgb}{0,0,0}

\newcommand{\gsec}[1]{{\hypersetup{linkcolor=blue}Sec.~\ref{#1}\hypersetup{linkcolor=red}}}

\newcommand{\geqn}[1]{(\ref{#1})}
\newcommand{\gfig}[1]{{\hypersetup{linkcolor=violet}Fig.~\ref{#1}\hypersetup{linkcolor=red}}}
\newcommand{\gtab}[1]{{\hypersetup{linkcolor=gesflanse}Table~\ref{#1}\hypersetup{linkcolor=red}}}

\begin{document}

\title{\textbf{\huge \titledef}} 
\author[1]{{\large Shao-Feng Ge}\footnote{gesf02@gmail.com}}
\author[1]{{\large Werner Rodejohann}\footnote{werner.rodejohann@mpi-hd.mpg.de}}
\affil[1]{\small Max-Planck-Institut f\"{u}r Kernphysik, Postfach 103980, 
69029 Heidelberg, Germany}
\date{\today}

\maketitle

\begin{abstract}
\noindent
We study the impact of the precision determination of oscillation parameters in the 
JUNO experiment on half-life predictions for neutrinoless double beta decay. 
We show that the solar neutrino mixing angle can be measured by JUNO with 
below $1\%$ uncertainty. This implies in particular that the minimal value
of the effective mass in the inverted mass ordering 
will be known essentially without uncertainty. We demonstrate that this 
reduces the range of half-life predictions in order to test this value 
by a factor of  two. The remaining uncertainty is caused by 
nuclear matrix elements. 
This has important consequences for future double beta decay experiments 
that aim at ruling out the inverted mass ordering or the Majorana nature of neutrinos.

\end{abstract}

\section{Introduction}

With neutrino mass and lepton mixing firmly established as facts, a minimal basis 
to describe those phenomena has been developed. In this ``3 Majorana neutrino paradigm'', 
three massive Majorana neutrinos are described by the neutrino 
mass matrix $m_\nu$: 
\begin{equation}\label{eq:mnu}
{\cal L} = \frac 12 \nu^T_\alpha (m_\nu)_{\alpha \beta} \, \nu_\beta \,.
\end{equation}
There are nine physical parameters in $m_\nu$, usually parametrized as 
3 masses, 3 mixing angles and 3 phases \cite{Agashe:2014kda}.  
Within this simple framework, 3 tasks are eminent: determining the parameters 
as precisely as possible; checking if the 
minimal description (3 Majorana neutrino paradigm) is correct; 
explaining the measured parameter values.

Yet to be determined experimentally are the neutrino mass ordering, 
the octant of the atmospheric mixing angle,  
the Dirac CP phase,  the absolute mass scale, whether neutrinos are Dirac or Majorana 
type fermions, and the Majorana CP phases if neutrinos are Majorana type. 
The first three play a role in neutrino oscillations, with which one can
determine in total 6 parameters, 
2 mass-squared differences, 3 angles and the Dirac phase. 
For the remaining parameters and properties 
non-oscillation experiments are inevitable. 
While our current information 
on the established oscillation 
parameters is already impressive \cite{Valle14}, further improvement is of course 
needed. For instance, this is necessary to rule out flavor symmetry
models (see e.g.\ \cite{Hanlon:2013ska}) or to 
check for new physics such as non-standard interactions, unitarity violation, 
long-range forces, etc. In this paper we address however another aspect of 
precision determination of neutrino oscillation parameters, namely its impact 
on neutrinoless double beta decay $(0 \nu 2 \beta)$ \cite{Rodejohann:2011mu,Pas:2015eia}. 

It is well-known that 
$0 \nu 2 \beta$ can in principle contribute to determining the Majorana or Dirac 
nature of neutrinos and also to the question of the neutrino mass ordering: in the 
standard paradigm the effective mass, on which the amplitude of the process depends, 
cannot vanish if the inverted mass ordering is realized. Therefore, if an oscillation 
experiment shows that neutrino masses have an inverted ordering, we 
know that the process must happen with a certain half-life $T_{1/2}^{\rm max}$. 
Not observing the decay 
with a half-life limit above $T_{1/2}^{\rm max}$ means that neutrinos are Dirac particles. 
In turn, not knowing the mass ordering and not observing the decay 
with a half-life limit above $T_{1/2}^{\rm max}$ rules out the inverted ordering in case 
neutrinos are Majorana particles\footnote{Of course, new physics in
  neutrinoless double 
beta decay such as light sterile neutrinos or TeV-scale 
left-right symmetric models could modify such statements, see 
Refs.\ \cite{Rodejohann:2011mu,Pas:2015eia} for a general discussion. We assume here that no 
contribution other than three massive Majorana neutrinos to double beta decay plays a role, thus we stay within the best motivated (standard) interpretation of the decay.}. 
The timescales of determining the mass ordering \cite{Qian:2015waa} and reaching 
half-life limits in the inverted ordering regime are 
comparable \cite{Schwingenheuer:2012zs} and subject to uncertainties, so
 both possibilities could happen. The effective mass value that future 
experiments need to reach in order to fulfill the two goals mentioned
above needs to be known as 
precisely as possible, since it enters the half-life 
quadratically.  If the experiment is dominated by background the
situation is even worse. 

As argued in \cite{Dueck:2011hu}, the upper limit on the minimal 
half-life in the inverted ordering depends strongly on the solar neutrino mixing angle 
and its current uncertainty introduces a sizable uncertainty in half-life predictions. 
This ``particle physics uncertainty'' is of the same order as the 
``nuclear physics uncertainty'', i.e.\ the notorious nuclear 
matrix elements. Both uncertainties should be reduced. 
In this paper we demonstrate that future precision data on the 
oscillation parameters essentially removes the particle physics uncertainty. 
Towards this end, we apply the NuPro \cite{nupro} package, written by one of us (SFG),  
to evaluate the precision with which the upcoming JUNO experiment \cite{JUNO} 
will determine the neutrino oscillation parameters and thus in particular 
the minimal effective mass in the inverted ordering. 
For demonstration, we will use JUNO as an example, we note that  
RENO-50 \cite{RENO50}, with a very similar configuration, will reach similar 
precision.

We demonstrate that such medium baseline reactor experiments 
 can measure the solar neutrino mixing angle with unprecedented precision 
of even less than 1\%, in addition to their main goal of 
measuring the neutrino mass ordering. We translate this precision in half-life ranges 
of neutrinoless double beta decay for various isotopes and nuclear matrix element 
calculations. This will help $0 \nu 2 \beta$ 
experiments to possibly rule out the Majorana nature and evaluate their requirements 
 to achieve this goal.  

 JUNO can not only determine the mass ordering, but also can 
measure the solar neutrino mixing angle with impressive precision at the same time. 
Thus, a single experiment can provide double beta decay
experiments with all necessary information 
for ruling out the Majorana nature. To be more precise, the minimal
value of the effective mass in the inverted mass ordering depends on 
the atmospheric mass-squared difference (very weakly on the solar one as well), $\theta_{12}$ and
$\theta_{13}$, i.e.\ the same set of parameters the
electron neutrino survival probability in reactor experiments
depends on. Thus, the latter have a direct correspondence to neutrinoless
double beta decay experiments (and in principle also to single beta
decay experiments). 

Of course, if JUNO determines that neutrinos enjoy a normal mass ordering, the 
interesting link to double beta decay is somewhat lost, though the uncertainty 
of the effective mass in the normal ordering will still be 
reduced significantly. In case JUNO would not be able to determine the mass ordering because 
of limited energy resolution, we show that the precision on the solar neutrino 
mixing angle is not affected.\\

This paper is organized as follows. In \gsec{sec:0nu2beta} we discuss 
$0 \nu 2 \beta$ with three Majorana neutrinos 
and how the ability of ruling out the inverted ordering or the Majorana nature 
is related to the uncertainty of neutrino oscillation parameters. 
Then, we evaluate the precision that can be achieved at JUNO
in \gsec{sec:JUNO}, which is used in \gsec{sec:IH} to derive the required half-life  
sensitivity that $0 \nu 2 \beta$ experiments need to provide. Our conclusion can be
found in \gsec{sec:conclusion}.

\section{Effective Neutrino Mass and Half-Life}
\label{sec:0nu2beta}

\begin{figure}[h!]
\centering
\includegraphics[height=0.32\textwidth,width=0.27\textwidth,angle=-90]{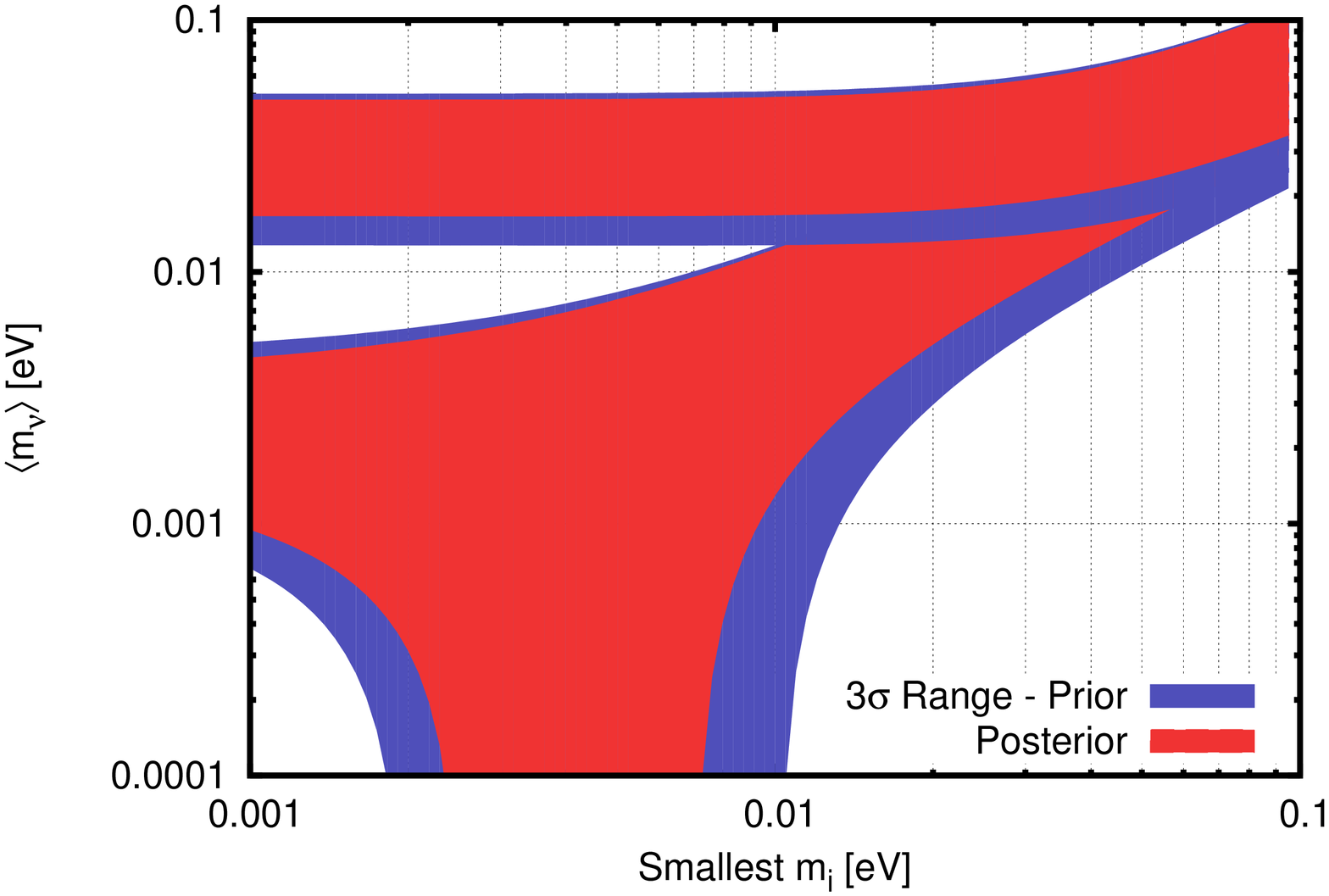}
\includegraphics[height=0.32\textwidth,width=0.27\textwidth,angle=-90]{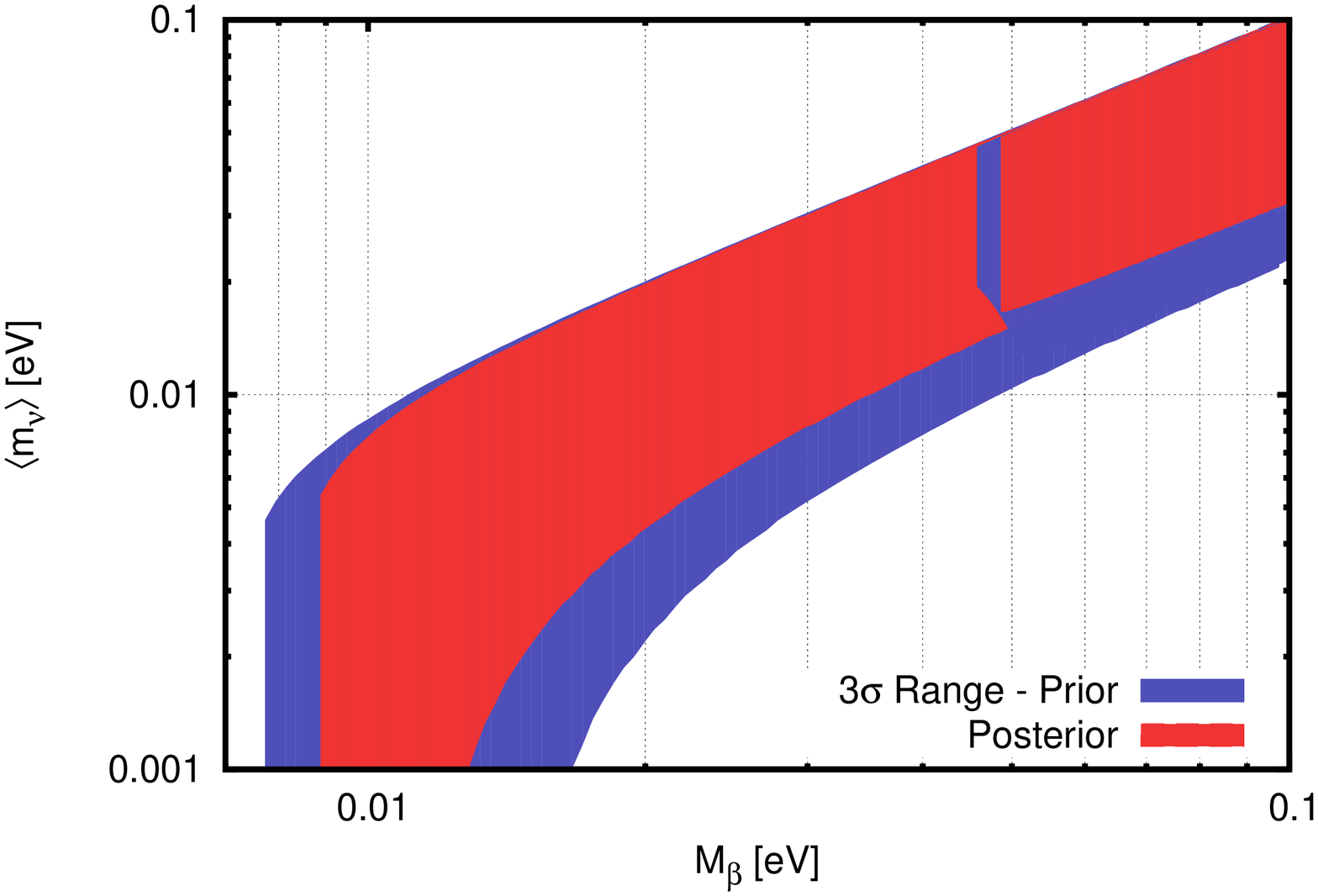}
\includegraphics[height=0.32\textwidth,width=0.27\textwidth,angle=-90]{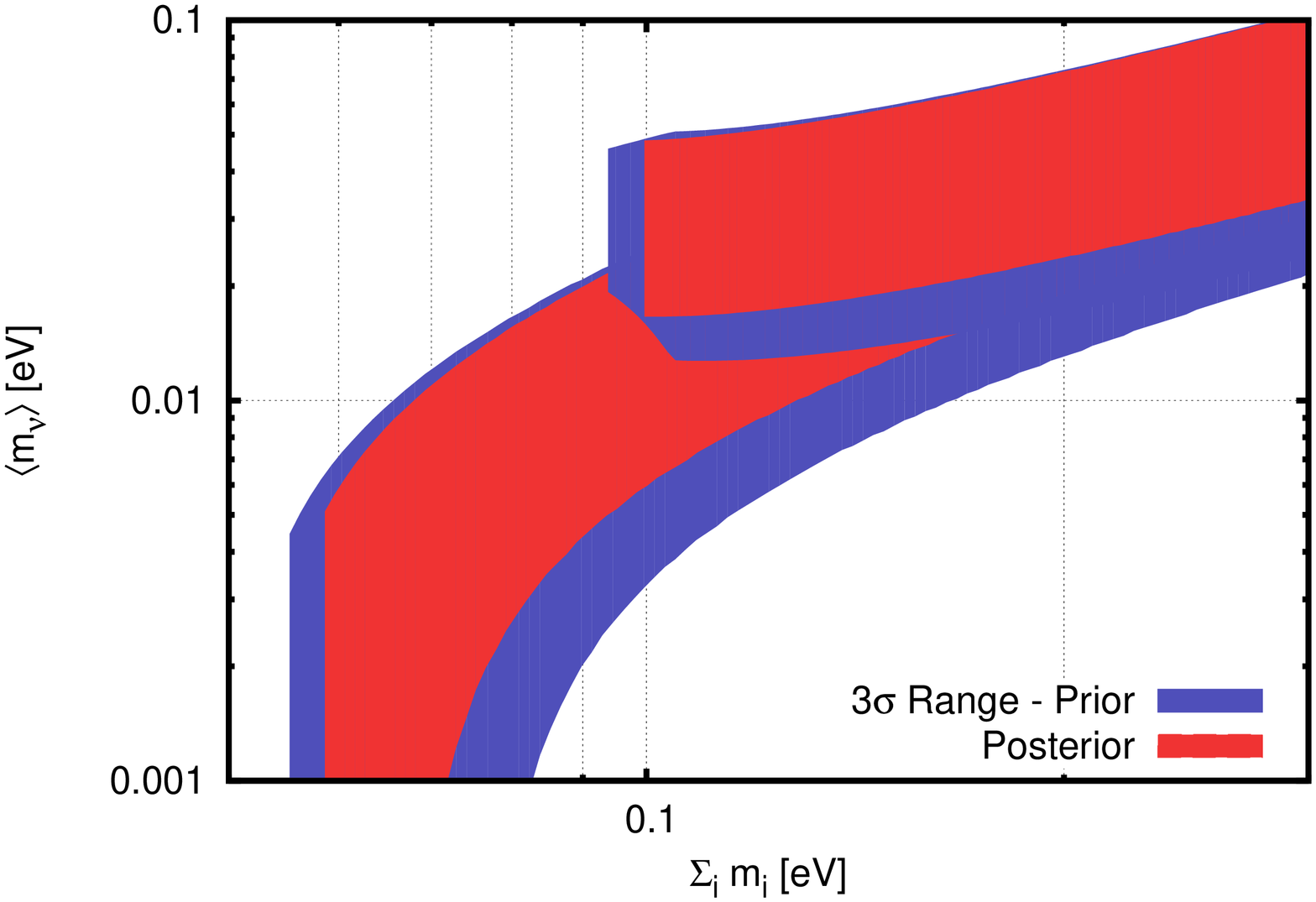}
\caption{The effective mass as a function of the smallest mass eigenvalue ($m_1$ for NO and $m_3$ for IO),
         $\beta$--decay mass $m_\beta=\sqrt{|U_{ei}|^2 m_i^2}$, and the sum of mass eigenvalues, within the $3\sigma$ range before
         (prior) and after (posterior) JUNO.}
\label{fig:obs}
\end{figure}

Neutrinoless Double Beta Decay ($0 \nu 2 \beta$) is fundamentally important to particle
physics as its observation implies lepton number violation, similar to the 
baryon number violation implied by an observation of proton decay. 
Several mechanisms for $0 \nu 2 \beta$ exist \cite{Rodejohann:2011mu,Pas:2015eia}. 
In the best motivated interpretation, $0 \nu 2 \beta$ is mediated by Majorana neutrinos 
with mixing observed in neutrino oscillation experiments. The decay half-life can be 
expressed as
\begin{equation}
  \left( T^{0 \nu}_{1/2} \right)^{-1}
=
  G^{0 \nu}
  |M^{0 \nu}|^2
  \langle m_\nu \rangle^2 \,,
\label{eq:Thalf}
\end{equation}
where $G^{0 \nu}$ is the well-known 
phase space factor and $M^{0 \nu}$ the nuclear matrix element. 
Neutrino mass and mixing enters through the effective electron 
neutrino mass $\langle m_\nu \rangle$ 
\begin{equation}
  \langle m_\nu \rangle
=
\left|
  c^2_s c^2_r m_1 
+ s^2_s c^2_r m_2 e^{i \alpha} 
+ s^2_r m_3 e^{i \beta}
\right| ,
\end{equation}
with $(c_\alpha, s_\alpha) \equiv (\cos \theta_\alpha, \sin \theta_\alpha)$. Here, we have
adopted the notation for mixing angles according to 
$\theta_a \equiv \theta_{23}$ denoting the atmospheric 
mixing angle, $\theta_r \equiv \theta_{13}$
the reactor mixing angle and $\theta_s \equiv \theta_{12}$ the solar mixing angle. For 
$0 \nu 2 \beta$, the only relevant mixing angles are $\theta_s$ and $\theta_r$. Apart from
these, the relevant parameters from the particle physics side involve the three neutrino
mass eigenvalues $m_i$ and the two Majorana CP phases $\alpha$ and $\beta$. 
Measured by neutrino oscillation experiments are the mixing angles and the 
two mass-squared  differences, $\Delta m^2_s \equiv m^2_2 - m^2_1$ and 
$\Delta m^2_a \equiv |m^2_3 - m^2_1|$. 
Note that for the mass-squared difference between $m^2_1$ and $m^2_3$ only the magnitude 
has been measured. 
The different sign of $\Delta m^2_a$ leads to quite different mass patterns, 
the normal ordering (NO) with $m_1 < m_2 < m_3$ and the inverted 
ordering (IO) with $m_3 < m_1 < m_2$. 
Fig.\ \ref{fig:obs} shows for both mass orderings the effective mass versus the smallest mass, 
as well as versus the neutrino mass parameters that are accessible in direct searches 
and cosmology. Both the current as well as future $3\sigma$ ranges, 
to be determined in the later part of the paper, are given.

Even with the absolute mass scale and mass ordering fixed, the effective 
electron neutrino mass 
$\langle m_\nu \rangle$ can still vary as a function of the two unknown Majorana CP phases
$\alpha$ and $\beta$. For IO, the maximal and minimal values are 
\begin{subequations}
\begin{eqnarray}
  \langle m_\nu \rangle^{\rm IO}_{\rm max}
& = &
  \sqrt{m^2_3 + \Delta m^2_a} c^2_s c^2_r
+ \sqrt{m^2_3 + \Delta m^2_a + \Delta m^2_s} s^2_s c^2_r
+ m_3 s^2_r \,,
\\
  \langle m_\nu \rangle^{\rm IO}_{\rm min}
& = &
  \sqrt{m^2_3 + \Delta m^2_a} c^2_s c^2_r
- \sqrt{m^2_3 + \Delta m^2_a + \Delta m^2_s} s^2_s c^2_r
- m_3 s^2_r \,.
\end{eqnarray}
\label{eq:IH-limits}
\end{subequations}
Since $m_3$ is the smallest mass, $s^2_r \ll c^2_r$, 
$c^2_s \simeq 2 s^2_s $ and $\Delta m^2_a \gg \Delta m^2_s$, 
the first term actually dominates over the second. 
As the two limits in \geqn{eq:IH-limits} increase with $m_3$, there 
is an universal lower limit \cite{Pascoli:2002xq} for IO when $m_3$ approaches 
zero (the lower limit $\langle m_\nu \rangle^{\rm IO}_{\rm min}$ can also easily be extracted 
within a geometrical picture \cite{Xing:2014yka}): 
\begin{equation}
  \langle m_\nu \rangle^{\rm IO}_{\rm min}
\rightarrow
  \sqrt{\Delta m^2_a} (c^2_s - s^2_s) c^2_r \,.
\label{eq:global-IHmin}
\end{equation}
This value can  put an upper bound on the half-life time $T^{0 \nu}_{1/2}$ through
\geqn{eq:Thalf}. To exclude the inverted ordering, 
it is necessary for $0 \nu 2 \beta$ experiments to go beyond this limit. How precisely 
do we know this limit? 
Currently, the atmospheric mass-squared difference $\Delta m^2_a$ and the reactor
mixing angle $\theta_r$ have been measured with precision at the few percent level. 
However, the solar mixing angle $\theta_s$ can contribute a significant 
uncertainty to $\langle m_\nu \rangle$ \cite{Dueck:2011hu}. 
\begin{figure}[b!]
\centering
\includegraphics[height=0.85\textwidth,angle=-90]{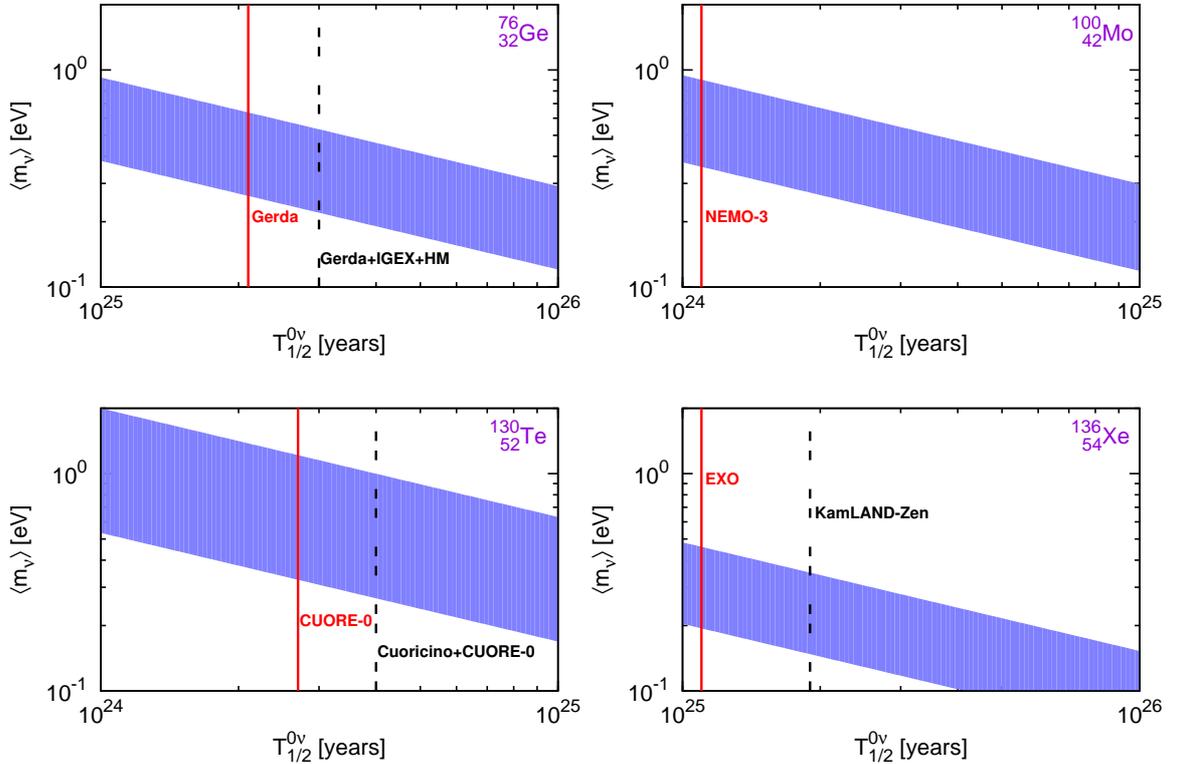}
\caption{Current limits on the effective mass  (90\% C.L.)  for some isotopes 
for the range of nuclear matrix elements given in Table \ref{tab:NME}. The vertical lines 
are current experimental half-life limits.}
\label{fig:current}
\end{figure}
\begin{figure}[h!]
\centering
\includegraphics[height=0.85\textwidth,width=0.5\textwidth,angle=-90]{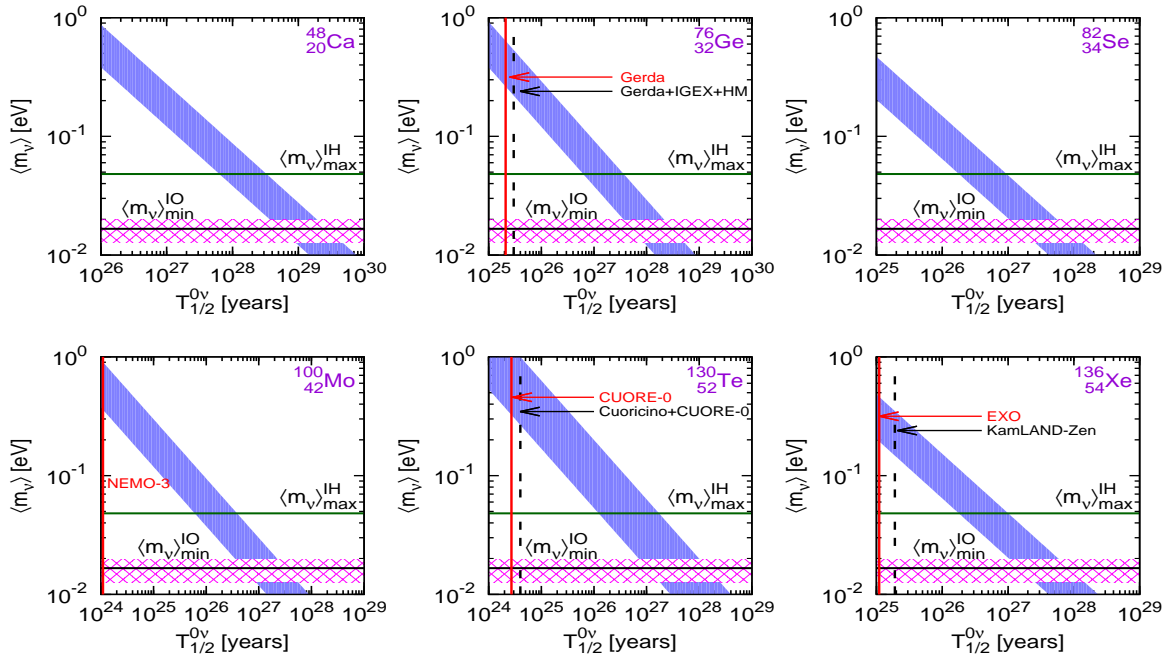}
\caption{Experimental constraints (90\% C.L.) on the effective mass for some isotopes 
for the range of nuclear matrix elements given in Table \ref{tab:NME}. The vertical lines 
are current experimental half-life limits, the horizontal band is the current 
range for the minimal value in the inverted ordering, the lower solid line is after JUNO. 
The upper solid line is for the half-life to enter the inverted ordering regime, 
$\langle m_\nu \rangle_{\rm max}^{\rm IH} = \sqrt{\Delta m^2_a} c^2_r$.}
\label{fig:mass}
\end{figure}

According to the latest global fit \cite{Valle14}, 
the $3 \sigma$ range of the solar mixing angle is 
$0.444 \leq \cos 2 \theta_s \leq 0.250$. This implies at $3 \sigma$ C.L.\ a 
total uncertainty of 
\begin{equation}
  \langle m_\nu \rangle^{\rm IO}_{\rm min} 
= 
  (0.0127 \ldots 0.0198) \mbox{ eV} \,,
\label{eq:main}
\end{equation}
which varies by a factor of $1.6$. Apart from $\langle m_\nu \rangle$, 
there is of course another source of uncertainty in predicting 
$T^{0 \nu}_{1/2}$, namely the nuclear matrix elements $M^{0 \nu}$. 
Their uncertainty, see \gtab{tab:NME}, is of similar magnitude as the 
one coming from the oscillation parameters, in particular 
$\theta_s$. Nuclear physics uncertainties will be addressed 
by the nuclear physics community. Here we focus on 
the particle physics uncertainty in $\langle m_\nu \rangle$, where progress 
is essentially guaranteed. 
In Fig.\ \ref{fig:current} we show, using the range of nuclear matrix elements 
from Table \ref{tab:NME}, for a subset of isotopes the current effective mass limits. 
The limits are 
\begin{equation}\label{eq:meff_lim}
\begin{array}{cc}
\langle m_\nu \rangle \le (0.22 \ldots 0.53) \mbox{ eV} & \mbox{ for } ^{76}{\rm Ge}\,,\\
\langle m_\nu \rangle \le (0.36 \ldots 0.90) \mbox{ eV} & \mbox{ for } ^{100}{\rm Mo}\,,\\
\langle m_\nu \rangle \le (0.27 \ldots 1.00) \mbox{ eV} & \mbox{ for } ^{130}{\rm Te}\,,\\
\langle m_\nu \rangle \le (0.15 \ldots 0.35) \mbox{ eV} & \mbox{ for } ^{136}{\rm Xe}\,,
\end{array}
\end{equation}
see also \cite{Guzowski:2015saa}. Note that these limits are not just
for the three light neutrinos from Eq.\ (\ref{eq:mnu}), but include in
principle possible sterile neutrinos up to masses slightly below the
Fermi scale, i.e.\ the constraint is actually on
$\sum\limits_{1,n}U_{ei}^2 m_i  $, where the sum goes over all neutrinos
below $m_n \simeq 50$ MeV. Note finally that the matrix elements
depend roughly quadratically on the axial coupling constant $g_A$, and
thus the effective mass limits in Eq.\ (\ref{eq:meff_lim}) would weaken
by $(1.27/g_A)^2$ in case $g_A$ would quench, i.e.\ become smaller as
the nuclear mass becomes larger \cite{Barea:2013bz}, see also 
\cite{Dell'Oro:2014yca}. 
The half-life
limits we obtain in what follows would become stronger by a factor
$(1.27/g_A)^4$. 
Fig.\ \ref{fig:mass} shows for a larger subset 
of isotopes the effective mass limits as a function of half-life limits, showing in particular 
the necessary 
half-lifes to rule out the inverted ordering, to be derived in Sec.\ \ref{sec:IH}. 
We also give the necessary half-lifes to enter the regime in which for the 
inverted ordering in case of vanishing smallest mass the effective mass should lie, 
$\langle m_\nu \rangle^{\rm IH}_{\rm max} = \sqrt{\Delta m^2_a} c^2_r$.

Note that the effective mass enters the half-life quadratically. 
If a $0 \nu 2 \beta$ experiment is dominated by background, the
half-life that an experiment can reach is proportional to \cite{Avignone:2005cs}
\begin{equation}\label{eq:bg}
  T^{0 \nu}_{1/2}
\propto
  a \times \epsilon \times \sqrt{\frac {M \times t}{B \times \Delta E}} \,,
\end{equation}
where $a$ is the isotopic abundance of the double beta emitter, $M$ the fiducial mass,
$t$ the measuring time, $\epsilon$ the detection efficiency, $\Delta E$ the energy
resolution around the peak, and $B$ the constant background rate in 
units of counts/keV/kg/yr. We see that a factor of 
2 uncertainty in the effective mass implies experimentally either a factor $4$ in 
half-life or a combined factor of $16$ in the experimental parameters
in the square root in 
\geqn{eq:bg}. For the current $3 \sigma$ range, a variation by a factor of $1.6$ in
$\langle m_\nu \rangle^{\rm IO}_{\rm min}$ corresponds to a factor of $2.56$
variation in the half-life, and thus a factor of $6$ in the
experimental parameters. 
 Obviously this is challenging, especially when 
it comes to estimating the necessary size of the detector and the runtime to 
reach the limit of excluding IO. 

In the next Section we will demonstrate how the precision on the effective 
mass can be improved at JUNO.

\section{Precision Measurement of Solar Mixing Angle at JUNO}
\label{sec:JUNO}

Reactor neutrinos are generated by nuclear reactions. Hence, their typical energy is in
the MeV range which is not enough to produce muon or tau leptons in the final state. 
The only accessible oscillation channel is the electron survival probability 
$P_{ee}$, which depends on only two mixing angles, $\theta_r$ 
and $\theta_s$, as well as two mass-squared differences:  
\begin{equation}
  P_{ee}
=
  1
- 4 c^4_r c^2_s s^2_s \sin^2 \Delta_{21}
- 4 c^2_s c^2_r s^2_r \sin^2 \Delta_{31}
- 4 s^2_s c^2_r s^2_r \sin^2 \Delta_{32} \,,
\label{eq:Pee}
\end{equation}
where $\Delta_{ij} \equiv (m^2_i - m^2_j) L / 4 E_\nu$. Note that, just as for the 
effective mass, the atmospheric mixing
angle $\theta_a$ and the Dirac CP phase $\delta_D$ are not involved. 
At short-baseline reactor experiments the dominant oscillation comes
from the last two terms which are modulated by the large mass-squared difference 
$\Delta m^2_{32} \approx \Delta m^2_{31} = \Delta m^2_a$. The amplitude $\sim c^2_r s^2_r$ 
is essentially independent of the solar mixing angle $\theta_s$, hence short-baseline
experiments can be used to measure the reactor mixing angle 
$\theta_r$ precisely. This has been and will be done by Daya-Bay, Double Chooz 
and RENO. 

\begin{figure}[t!]
\centering
\includegraphics[height=0.505\textwidth,angle=-90]{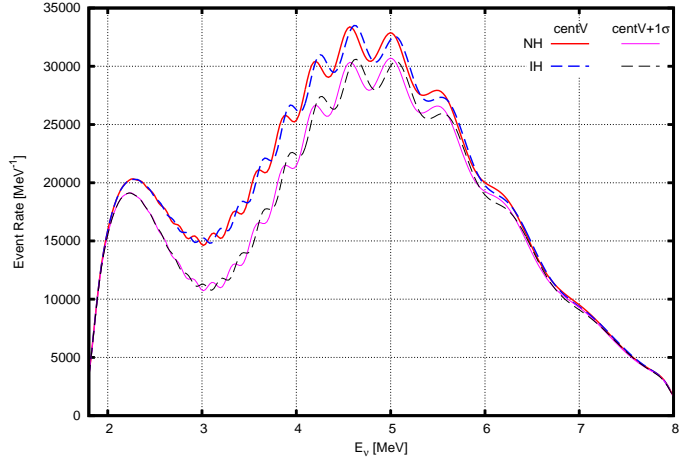}
\caption{The event rate of reactor neutrinos to be observed at
  JUNO. Two values of the solar neutrino mixing angle were used,
  $\sin^2 \theta_s = 0.323$ (thick lines) and $\sin^2 \theta_s = 0.339$ (thin lines).}
\label{fig:rate}
\end{figure}

On the other hand, at longer baseline the contribution from the second term in 
\geqn{eq:Pee} becomes dominant. 
On top of this slow oscillation, many fast oscillations occur. 
Note that the last two terms in \geqn{eq:Pee} are not exactly the same. The difference 
between $\Delta_{31}$ and $\Delta_{32}$ is roughly $\Delta m^2_s / \Delta m^2_a \approx 3\%$.
In other words, there are two fast oscillation modes with slight difference. 
This can be used to determine the neutrino 
mass ordering \cite{Petcov:2001sy,Choubey:2003qx}. 
Depending on the neutrino mass ordering  
one frequency is faster than the other. To make the picture clear,
we can formulate the oscillation \geqn{eq:Pee} in terms of $\Delta m^2_s$ and $\Delta m^2_a$,
\begin{eqnarray}
  P_{ee}
& = &
  1
- 4 c^4_r c^2_s s^2_s \sin^2 \Delta_{21}
- 4 c^2_r s^2_r \sin^2 |\Delta_{31}|
\nonumber
\\
& &
\phantom{1}
- 4 s^2_s c^2_r s^2_r \sin^2 \Delta_{21} \cos \left( 2 |\Delta_{31}| \right)
\pm 2 s^2_s c^2_r s^2_s \sin \left(2 \Delta_{21} \right) \sin \left(2 |\Delta_{31}| \right) ,
\end{eqnarray}
where $\pm$ corresponds to NO (+) and IO ($-$) respectively \cite{Ge12}. The difference in 
oscillation probability is equivalent to a relative phase shift between the 
two fast frequencies. Since the relative phase difference is only about 
$\Delta m^2_s / \Delta m^2_a \approx 3\%$, 
it is necessary for the energy resolution to be better than $3\%$. 
By measuring the relative phase shift
between the two fast frequencies, the neutrino mass ordering can be determined 
\cite{Learned:2006wy, Zhan:2008id, Zhan:2009rs}. This corresponds to measuring the atmospheric
mass-squared difference $\Delta m^2_a$ with precision around
$\Delta m^2_s / \Delta m^2_a \approx 3\%$, so that the tiny difference between $\Delta m^2_{31}$
and $\Delta m^2_{32}$ can be seen.
To illustrate the picture, we show the event rate in \gfig{fig:rate}.
In the current study we focus on the JUNO configuration \cite{JUNO} with a 20kt liquid scintillator
detector 52 km away from two reactor complexes with 36 GW total thermal power. The effective
runtime is roughly $4.9$ years (normal run time of 6 years with 300 effective days per
year) and a detector energy resolution $3\%/\sqrt{E\,(\mbox{MeV})}$ as benchmark.

One side product from this medium baseline reactor neutrino experiment is that the low 
frequency oscillation due to the second term in \geqn{eq:Pee}, 
which is essentially modulated
by $4 c^2_s s^2_s$, can be used to measure the solar mixing angle $\theta_s$ precisely 
\cite{Batygov:2008ku,Ge12}. 
This constraint mainly comes from the total event rate as shown 
in \gfig{fig:rate}. 
Consequently, it will not be affected much by energy resolution, in contrary 
to the determination of the neutrino mass ordering. 
A rough estimate gives $1 \times 10^5$ events to be observed,  
with a statistically fluctuation of $3 \times 10^{-3}$. The actual uncertainty should 
be even further suppressed since not only the total event rate is measured, but a whole
spectrum across the whole energy range. 
The position of the main peak in \gfig{fig:rate} can be used to constrain the corresponding 
oscillation frequency $\Delta m^2_s$. Neither of them is sensitive 
to the energy resolution that the detector can finally achieve. 
Therefore a precise measurement of the solar mixing angle $\theta_s$ and the solar oscillation 
frequency $\Delta m^2_s$ is guaranteed.
In \gfig{fig:rate}, we also show the dependence on the solar mixing angle by plotting
the event rates with central value $\sin^2 \theta_s = 0.323$ and $1 \sigma$ upper
limit $\sin^2 \theta_s = 0.339$, respectively. 

We note that there is a study showing that the 5 MeV reactor anomaly
can introduce a small shift in the best-fit of $\theta_s$ \cite{Ciuffoli:2015dsa}. 
This anomaly is a bump in the ratio of measured and observed neutrino
fluxes around energies of 5 MeV, as observed at Daya Bay \cite{DayaBay-anomaly}, 
Double CHOOZ \cite{DoubleCHOOZ-anomaly}, and
RENO \cite{RENO-anomaly-1,RENO-anomaly-2}. 
For a measurement of $\theta_s$ the anomaly can change the total event
rate which is mainly controlled
by $4 c^2_s s^2_s$ as we explained above. Consequently, the best fit of $\theta_s$ 
can be shifted away from its true value, although its uncertainty would not be 
affected much. This difficulty could in principle be 
overcome by an extra near detector. By comparing the event rates measured at near
and far detectors, we can extract the oscillation probability between the two detectors
that is purely due to the conventional three-neutrino oscillation.
In addition, fitting the data with a parametrizable flux depletion/excess, such
as gaussian-like shape as implemented in \cite{Ciuffoli:2015dsa}, can also 
correct the best fit-value. 

\begin{figure}[t!]
\centering
\includegraphics[height=0.505\textwidth,angle=-90]{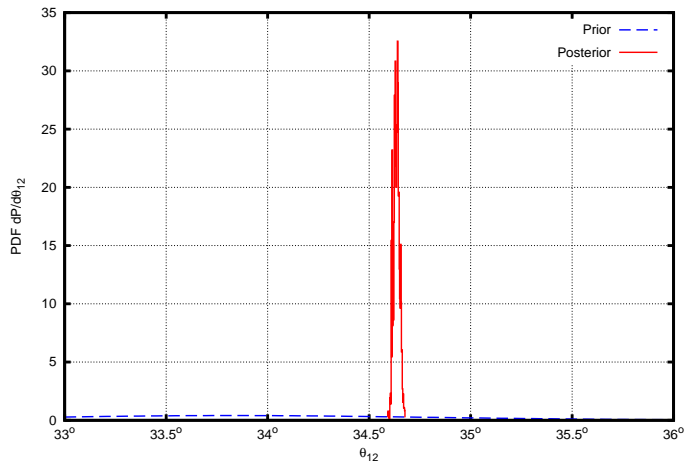}
\caption{Prior (input) and posterior (output) distributions of the solar mixing angle $\theta_s$ at JUNO.}
\label{fig:ts}
\end{figure}
In \gfig{fig:ts} we show the prior (input) and posterior (output) distributions 
of the solar mixing angle at JUNO, simulated with NuPro \cite{nupro}. In total, the whole range 
from $1.8$ MeV to 8 MeV is divided into $400$ bins equally. Since the event rate in each bin is 
more that $20$ events, the statistical fluctuation can be approximated as gaussian distribution.
The $\chi^2$-function is defined as 
\begin{equation}
  \chi^2
\equiv
  \sum_i \frac {(\overline N_i - f N_i)^2}{\overline N_i} 
+ \left[ \hspace{-0.5mm} \frac {\Delta m^2_s - \overline{\Delta m^2_s}}{\delta(\Delta m^2_s)} \right]^2 
\hspace{-2mm}
+ \left[ \hspace{-0.5mm} \frac {\Delta m^2_a - \overline{\Delta m^2_a}}{\delta(\Delta m^2_a)} \right]^2
\hspace{-2mm}
+ \left[ \hspace{-0.5mm} \frac {\sin^2 2 \theta_r - \overline{\sin^2 2 \theta_r}}{\delta(\sin^2 2 \theta_r)} \right]^2
\hspace{-2mm}
+ \left[ \hspace{-0.5mm} \frac {\sin^2 \theta_s - \overline{\sin^2 \theta_s}}{\delta(\sin^2 \theta_s)} \right]^2
\hspace{-2mm}
+ \frac {(f - 1)^2}{0.01^2} \,,
\label{eq:chi2}
\end{equation}
with input values,
\begin{subequations}
\begin{eqnarray}
  \Delta m^2_s = (7.6 \pm 0.2) \times 10^{-5} \,\mbox{eV}^2 \,,
&&
  \Delta m^2_a = (2.4 \pm 0.1) \times 10^{-3} \,\mbox{eV}^2 \,,
\\
  \sin^2 2 \theta_r = 0.089 \pm 0.005 \,,
&&
  \sin^2 \theta_s = 0.323 \pm 0.016 \,,
\end{eqnarray}
\label{eq:inputs}
\end{subequations}
according to the latest global fit \cite{Valle14} and 
Daya Bay measurement on $\theta_r$ \cite{dayabay}.
In addition, the flux has an overall 1\% uncertainty parametrized as an overall normalization $f$.
Instead of minimization, the $\chi^2$-function \geqn{eq:chi2}
is used to sample the oscillation parameters, two mass-squared differences $\Delta m^2_s$ and 
$\Delta m^2_a$ as well as the two mixing angles $\theta_r$ and $\theta_s$, with the Bayesian Nested
Sampling algorithm \cite{skilling2006}. Since there is no measurement of the two Majorana CP phases,
they are randomly sampled between $0$ and $2 \pi$. The sampled points are then analyzed to get the best fit 
values, uncertainties, and ranges of various functions of oscillation parameters simultaneously. 
In this way, the maginalization is carried out automatically.

It turns out that JUNO can significantly improve the uncertainty
on $\theta_s$ from $\Delta[\sin^2 \theta_s] = 0.016$ down to 
$2.4 \times 10^{-3}$ which is consistent with the rough estimation based on total event numbers. 
Correspondingly, the uncertainty 
$\Delta(\theta_s)$ is as small as $0.15^\circ$, roughly $4 \times 10^{-3}$ of $\theta_s$ itself.
At this precision, we can claim that the solar mixing angle $\theta_s$ as well as $\cos 2 \theta_s$ 
are precisely known. In addition, the atmospheric mass-squared difference $\Delta m^2_a$ is
measured with precision $\Delta[\Delta m^2_a]/\Delta m^2_a \simeq 6 \times 10^{-3}$. In other words,
there is almost no contribution from the neutrino mixing sector to the global lower limit
of $\langle m_\nu \rangle^{\rm IO}_{\rm min}$ as defined in \geqn{eq:global-IHmin}. This will 
be quantified in the next Section.

\section{Implications for Double Beta Decay}
\label{sec:IH}

With the solar mixing angle $\theta_s$ and the atmospheric mass-squared difference $\Delta m^2_a$
precisely measured by huge statistics and excellent energy resolution, respectively, the global lower
limit on the effective electron neutrino mass \geqn{eq:global-IHmin} is actually fixed.
\begin{figure}[t!]
\centering
\includegraphics[height=0.5\textwidth,angle=-90]{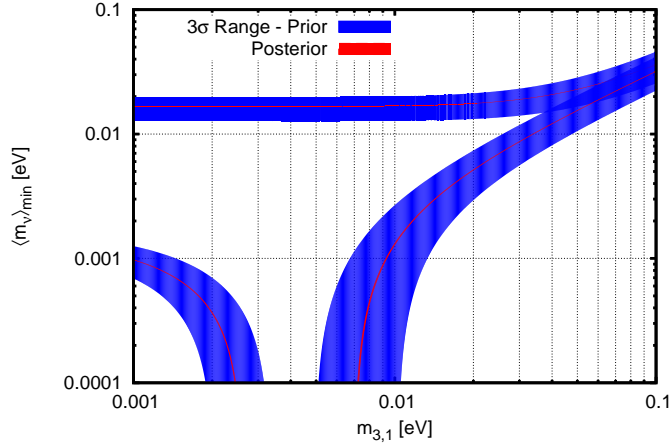}
\caption{The $3 \sigma$ range of $\langle m_\nu \rangle_{\rm min}$ as a function of
         the lightest neutrino mass, $m_1$ for NO and $m_3$ for IO. All oscillation parameters, including the
         solar and reactor mixing angles, $\theta_s$ and $\theta_r$, as well as the atmospheric 
         and solar mass-squared differences, $\Delta m^2_a$ and $\Delta m^2_s$, are varied 
         in their corresponding $3 \sigma$ ranges.}
\label{fig:minMee}
\end{figure}
In \gfig{fig:minMee}, we show the predicted $\langle m_\nu \rangle^{\rm IO}_{\rm min}$ 
as a function of the smallest neutrino mass eigenvalue, in this case $m_3$, with all other 
parameters, including the solar and reactor mixing angles, $\theta_s$ and $\theta_r$, as well 
as the atmospheric and solar mass-squared differences, $\Delta m^2_a$ and $\Delta m^2_s$, 
varied in their corresponding present and future $3 \sigma$ ranges\footnote{For 
completeness, we also give in \gfig{fig:minMee} 
the corresponding improvement for the minimal value 
in the normal ordering.}. 
At any value of $m_3$, the uncertainty of the
lower limit $\langle m_\nu \rangle^{\rm IO}_{\rm min}$ is compressed to negligible size. 
Without JUNO, the value of $\langle m_\nu \rangle^{\rm IO}_{\rm min}$
is unknown within a  
factor of roughly $1.6$ at $3 \sigma$ confidence level. After JUNO
is included, it drops to the level of $2 \times 10^{-3}$, which is a huge improvement. 
It is safe to claim that $\langle m_\nu \rangle^{\rm IO}_{\rm min}$ can be predicted
almost precisely. For NO, the improvement is also significant, as depicted in \gfig{fig:minMee}. 
We also show in \gfig{fig:obs} the effective mass versus the smallest 
mass eigenvalue, as well as versus the neutrino mass parameters that are accessible in direct searches 
and cosmology, for both mass orderings. The improvement from the current $3\sigma$ ranges is obvious. 

\begin{figure}[t]
\centering
\includegraphics[height=0.95\textwidth,angle=-90]{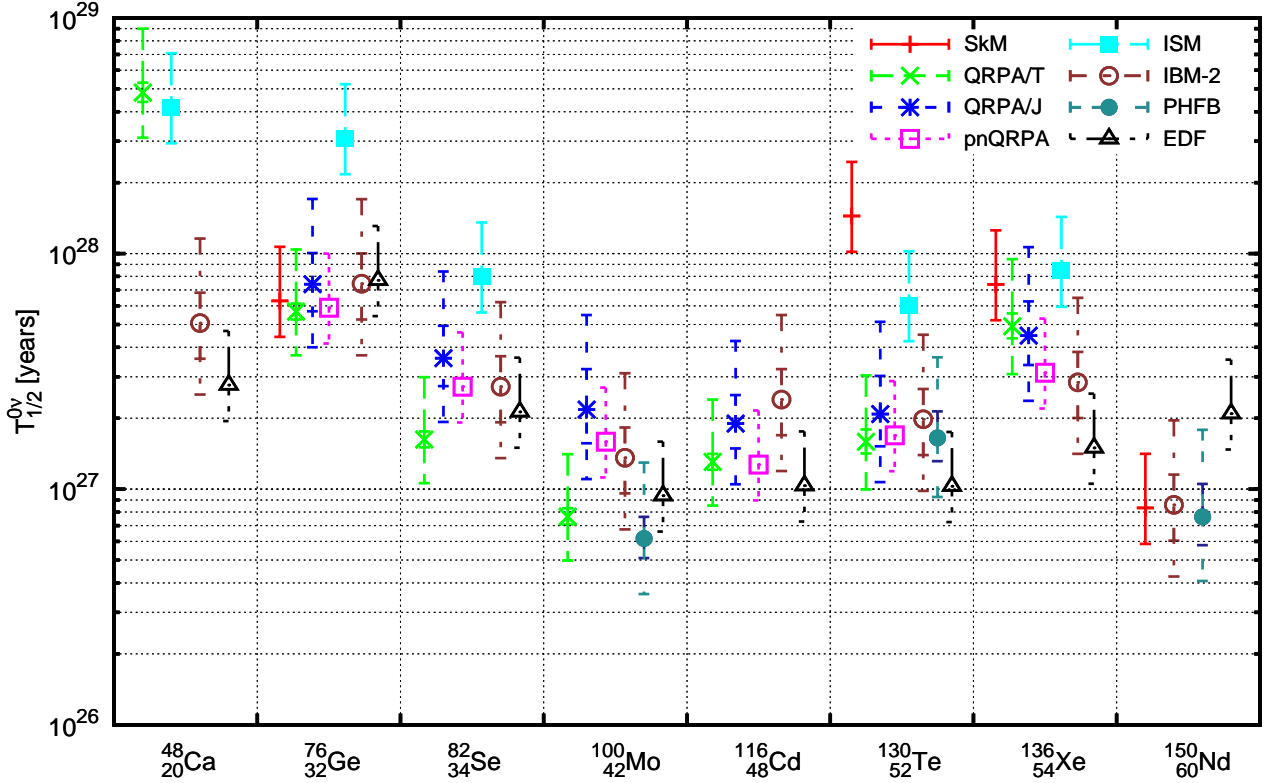}
\caption{The half-life $T^{0 \nu}_{1/2}$  of neutrinoless double beta decay 
        ($0 \nu 2 \beta$) to rule out the inverted ordering  
and its uncertainties. If one error bar is displayed, e.g.\ for SkM, 
 the range is
according to the current 3$\sigma$ uncertainty in the oscillation
parameters, which will be reduced to a single point after JUNO. 
If two error bars are displayed, e.g.\ for IBM-2, the nuclear 
matrix elements have a range, see Table \ref{tab:NME}. The larger
error bar is then the combined uncertainty, from the matrix elements
and the oscillation parameters in $\langle m \rangle_{\rm min}^{\rm
  IO}$. 
}
\label{fig:Thalf}
\end{figure}

\gfig{fig:Thalf} shows the necessary half-lifes to rule out the 
inverted ordering for the isotopes 
$^{48}$Ca, $^{76}$Ge, $^{82}$Se, $^{100}$Mo, $^{116}$Cd, 
$^{130}$Te, $^{136}$Xe and $^{150}$Nd. 
We have used a compilation of nuclear matrix elements which is 
summarized in Table \ref{tab:NME}. Also in this plot the improvement from the current 
range to the future situation is obvious. The shorter error bar corresponds to future 
uncertainty, which solely comes from the nuclear matrix elements. 
For instance, the range to go below $\langle m \rangle_{\rm min}^{\rm IO}$ for $^{76}$Ge 
is currently $T^{0 \nu}_{1/2} = (3.7 \ldots 52.3)\times 10^{27}$ yrs, while after JUNO it would be 
$T^{0 \nu}_{1/2} = (5.3\ldots 31) \times 10^{27}$ yrs, the range is
smaller by a factor 2.4. For $^{136}$Xe the range is currently 
$T^{0 \nu}_{1/2} = (1.1\ldots 14.3) \times 10^{27}$ yrs, which would
improve to  $T^{0 \nu}_{1/2} = (1.5 \ldots 8.4) \times 10^{27}$ yrs. 
See also Fig.\ \ref{fig:mass} for the necessary half-lifes to rule out (and reach, where the 
experimental improvement is of minor importance) 
the inverted ordering. For convenience, the ranges of half-lifes have 
been summarized in \gtab{tab:Thalf}. 

\section{Conclusion}
\label{sec:conclusion}
The next generation of neutrino oscillation experiments will be able
to provide mixing parameter determinations with remarkable
precision. This has impact on a variety of aspects. Here we have
studied inasmuch future reactor neutrino experiments can help in
planing and interpreting future searches for neutrinoless double beta
decay. The minimal value of the effective mass in the inverted
ordering is currently uncertain 
by a factor of almost 2, comparable to the nuclear matrix element uncertainty. 
Facilities like JUNO can determine the minimal value of the effective 
mass with essentially no uncertainty, which fixes the half-life values 
corresponding to those extreme values, up to the nuclear matrix
element uncertainties. The total uncertainty is therefore reduced by a
factor 2, leaving further improvement to the nuclear physics community.

\section*{Acknowledgements}
This work is partially supported by the Max Planck Society in the project MANITOP (WR).


\bibliographystyle{hunsrt}
\bibliography{nuless_IH_ts}
\nocite{*}

\begin{table}[h]
\setlength{\tabcolsep}{1.7mm}
\centering
\setlength\extrarowheight{1mm}
\small
\begin{tabular}{r|cccccccc|c}
  & $\rm SkM^{\cite{Mustonen:2013zu}}$ & $\rm QRPA/T^{\cite{Simkovic:2013qiy}}$ & $\rm QRPA/J^{\cite{Civitarese:2009zza}}$ & $\rm pnQRPA^{\cite{Hyvarinen:2015bda}}$ & $\rm ISM^{\cite{Menendez:2008jp}}$ & IBM-2${}^{\cite{Barea:2015kwa}}$ & PHFB$^{\cite{Rath:2013fma}}$ & EDF$^{\cite{Rodriguez:2010mn}}$ & $G^{0 \nu}{}^{\cite{Kotila:2012zza}}$ \\
\hline
${}^{48}_{20}\rm Ca$ & -- & 0.541$\ldots$0.594 & -- & -- & 0.61 & $1.75 \times (1 \pm 0.16)$ & -- & 2.37 & 23.29 \\[1mm] 
${}^{76}_{32}\rm Ge$ & 5.09 & 5.157$\ldots$5.571 & 4.029$\ldots$5.355 & 5.26 & 2.30 & $4.68 \times (1 \pm 0.16)$ & -- & 4.60 & 2.218 \\[1mm]
${}^{82}_{34}\rm Se$ & -- & 4.642$\ldots$5.018 & 2.771$\ldots$3.722 & 3.73 & 2.18 & $3.73 \times (1 \pm 0.16)$ & -- & 4.22 & 9.537 \\[1mm]
${}^{100}_{\phantom{1} 42} \rm Mo$ & -- & 5.402$\ldots$5.850 & 2.737$\ldots$3.931 & 3.90 & -- & $4.22 \times (1 \pm 0.16)$ & $6.26 \pm 0.63$ & 5.08 & 14.94 \\[1mm]
${}^{116}_{\phantom{1} 48} \rm Cd$ & -- & 4.040$\ldots$4.367 & 3.034$\ldots$3.935 & 4.26 & -- & $3.10 \times (1 \pm 0.16)$ & -- & 4.72 & 15.68 \\[1mm]
${}^{130}_{\phantom{1} 52} \rm Te$ & 1.37  & 3.888$\ldots$4.373 & 2.993$\ldots$4.221 & 4.00 & 2.12 & $3.70 \times (1 \pm 0.16)$ & $4.05 \pm 0.49$ & 5.13 & 13.35 \\[1mm]
${}^{136}_{\phantom{1} 54} \rm Xe$ & 1.89 & 2.177$\ldots$2.460 & 2.053$\ldots$2.802 & 2.91 & 1.77 & $3.05 \times (1 \pm 0.16)$ & -- & 4.20 & 13.69 \\[1mm]
${}^{150}_{\phantom{1} 60} \rm Nd$ & 2.71 & -- & -- & -- & -- & $2.67 \times (1 \pm 0.16)$ & $2.83 \pm 0.42$ & 1.71 & 59.16 
\end{tabular}
\caption{Theoretical predictions for the nuclear matrix elements and
  the  phase space factor in units of $10^{-26} \mbox{ yr}^{-1} \mbox{ eV}^{-2}$.}
\label{tab:NME}
\end{table}

\begin{table}[h!]
\setlength{\tabcolsep}{5.mm}
\centering
\setlength\extrarowheight{2mm}
\begin{tabular}{c|cc|c}
\multirow{2}{*}{$T_{1/2} \,\, [\mbox{yrs}]$}     & w/o JUNO & with JUNO     & \multirow{2}{*}{NME} \\
                                  & $\langle m_\nu \rangle _{\rm
                                    min}^{\rm IO} = (0.0127 \ldots
                                  0.0198) \mbox{ eV}$ & $\langle m_\nu
                                  \rangle _{\rm min}^{\rm IO} = 0.0166 \mbox{ eV}$ \\
\hline \hline
\multirow{2}{*}{${}^{48}\rm Ca$}  & $(1.95 \ldots 4.69) \times 10^{27}$ & $2.76 \times 10^{27}$ & $M^{0 \nu}_{\rm max} = 2.37$ \\
                                  & $(3.74 \ldots 9.01) \times 10^{28}$ & $5.30 \times 10^{28}$ & $M^{0 \nu}_{\rm min} = 0.55$ \\
\hline
\multirow{2}{*}{${}^{76}\rm Ge$}  & $(3.70 \ldots 8.92) \times 10^{27}$ & $5.25 \times 10^{27}$ & $M^{0 \nu}_{\rm max} = 5.57$ \\
                                  & $(2.17 \ldots 5.23) \times 10^{28}$ & $3.08 \times 10^{28}$ & $M^{0 \nu}_{\rm min} = 2.30$ \\
\hline
\multirow{2}{*}{${}^{82}\rm Se$}  & $(1.06 \ldots 2.56) \times 10^{27}$ & $1.51 \times 10^{27}$ & $M^{0 \nu}_{\rm max} = 5.02$ \\
                                  & $(5.62 \ldots 13.5) \times 10^{27}$ & $7.98 \times 10^{27}$ & $M^{0 \nu}_{\rm min} = 2.18$ \\
\hline
\multirow{2}{*}{${}^{100}\rm Mo$} & $(3.59 \ldots 8.65) \times 10^{26}$ & $5.10 \times 10^{26}$ & $M^{0 \nu}_{\rm max} = 6.89$ \\
                                  & $(2.28 \ldots 5.48) \times 10^{27}$ & $3.23 \times 10^{27}$ & $M^{0 \nu}_{\rm min} = 2.74$ \\
\hline
\multirow{2}{*}{${}^{116}\rm Cd$} & $(7.29 \ldots 17.6) \times 10^{26}$ & $1.04 \times 10^{27}$ & $M^{0 \nu}_{\rm max} = 4.72$ \\
                                  & $(2.28 \ldots 5.48) \times 10^{27}$ & $3.23 \times 10^{27}$ & $M^{0 \nu}_{\rm min} = 2.67$ \\
\hline
\multirow{2}{*}{${}^{130}\rm Te$} & $(7.25 \ldots 17.5) \times 10^{26}$ & $1.03 \times 10^{27}$ & $M^{0 \nu}_{\rm max} = 5.13$ \\
                                  & $(1.02 \ldots 2.45) \times 10^{28}$ & $1.44 \times 10^{28}$ & $M^{0 \nu}_{\rm min} = 1.37$ \\
\hline
\multirow{2}{*}{${}^{136}\rm Xe$} & $(1.06 \ldots 2.54) \times 10^{27}$ & $1.50 \times 10^{27}$ & $M^{0 \nu}_{\rm max} = 4.20$ \\
                                  & $(5.94 \ldots 14.3) \times 10^{27}$ & $8.43 \times 10^{27}$ & $M^{0 \nu}_{\rm min} = 1.77$ \\
\hline
\multirow{2}{*}{${}^{150}\rm Nd$} & $(4.08 \ldots 9.82) \times
10^{26}$ & $5.78 \times 10^{26}$ & $M^{0 \nu}_{\rm max} = 3.25$ \\
                                  & $(1.47 \ldots 3.55) \times 10^{27}$ & $2.09 \times 10^{27}$ & $M^{0 \nu}_{\rm min} = 1.71$ \\
\end{tabular}
\caption{Half-life ranges to rule out the inverted ordering 
before (after) JUNO. For each isotope, the
  first row corresponds to the range (or value) for the largest matrix
  element $M^{0 \nu}_{\rm max}$, while the second row is the range (value)
  for the smallest matrix element $M^{0 \nu}_{\rm min}$. 
Before JUNO (w/o), 
$\langle m_\nu \rangle_{\rm min}^{\rm IO}$ varies due to the
particle physics uncertainty which essentially vanishes after JUNO.}
\label{tab:Thalf}
\end{table}

\end{document}